\documentclass[camera]{gji}
\usepackage{timet}
\usepackage{graphicx}
\usepackage{amssymb}
\usepackage{amsmath}
\usepackage{psfrag}
\usepackage{multirow}
\usepackage{multicol}
\usepackage{wasysym}
\usepackage{url}
\usepackage{color}

\title[Upward-going cosmic muons]{Effects of upward-going cosmic muons on density radiography of volcanoes}

\author[K. Jourde  et al.]
{Kevin Jourde$^1$, Dominique Gibert$^{1,3}$, Jacques Marteau$^2$, Jean de Bremond d'Ars$^3$,
\and
Serge Gardien$^2$, Claude Girerd$^2$, Jean-Christophe Ianigro$^2$ \& Daniele Carbone$^4$ \\
$^1$ Institut de Physique du Globe de Paris, Sorbonne Paris Cit\'e, Univ Paris Diderot, UMR 7154 CNRS, Paris, France \\
$^2$ Institut de Physique Nucl\'eaire de Lyon, Univ Claude Bernard, UMR 5822 CNRS, Lyon, France. \\
$^3$ G\'eosciences Rennes, Univ Rennes 1, UMR 6118 CNRS, Rennes, France. \\
$^4$ Istituto Nazionale di Geofisica e Vulcanologia – Sezione di Catania, Catania, Italy.}

\date{}
\pagerange{\pageref{firstpage}--\pageref{lastpage}}
\volume{}
\pubyear{}

\begin{document}

\label{firstpage}
\maketitle

\begin{summary}

Muon tomography aims at deriving the density structure of geological bodies from their screening attenuation produced on the natural cosmic muons flux. Because of their open-sky exposure, muons telescopes are subject to noise fluxes with large intensities relative to the tiny flux of interest. A recognized source of noise flux comes from fake tracks caused by particles that fortuitously trigger the telescope detectors at the same time. Such a flux may be reduced by using multiple-detector telescopes so that fortuitous events become very unlikely. In the present study, we report on a different type of noise flux caused by upward-going muons crossing the detectors from the rear side. We describe field experiments on La Soufri\`ere of Guadeloupe and Mount Etna, and give details on the high-resolution clocking system and the statistical procedure necessary to detect upward-going muons. We analyse several data sets acquired either in calibration or in volcano tomography situation. All data sets are shown clearly biased by upward-going noise flux whose intensity may amount to $50\%$ of the measured total flux in given directions. Biases produced on density radiographies by this kind of flux are quantified and correction procedures are detailed. Examples for La Soufri\`ere and Mount Etna are given.

\end{summary}


\section{Introduction}

Density radiography with cosmic muons aims at determining the density of geological bodies by using the attenuation of the flux of cosmic muons caused by the screening effect of the rock volume to probe (Alvarez et al. 1970, Nagamine 1995; Nagamine et al. 1995; Tanaka et al. 2001). This new imaging technique presently deserves much interest because it allows to reconstruct the density distribution through a straight-ray tomography approach. Another great advantage of muon radiography is its ability to remotely tomography inaccessible targets like active unapproachable volcanoes (Tanaka et al. 2005; Gibert et al. 2010; Lesparre et al. 2012c).

The most recent applications of muon density radiography of volcanoes concern the monitoring of density variations related to either magma degassing in volcano chimneys (Tanaka et al. 2009; Shinohara and Tanaka 2012) or liquid/vapour transition in hydrothermal systems. These applications put severe constrains on the data quality and necessitate higher signal-to-noise ratio than demanded by more classical structural imaging where the density contrasts to be detected are generally high (e.g. Tanaka et al. 2001; Lesparre et al. 2012c; Portal et al. 2012; Carloganu et al. 2013). Indeed, imaging small-density contrasts implies detecting tiny variations in the flux of muon crossing the object of interest (Nagamine 2003; Lesparre et al. 2010), and any perturbing noise may definitely blur the relevant information.

In practice, muon radiography is done with telescopes that count the particles arriving from a given direction during a known period of time that may vary from days to months depending on the rock thickness and on the density resolution to reach (Lesparre et al. 2010). A main source of noise polluting the particle counting comes from fake tracks produced by particles that simultaneously hit the telescope detectors in such a way that their impacts could be interpreted as those produced by a single particle crossing the detectors (Lecomte 1963). Such a noise can be almost eliminated by using three or more detectors together with high-resolution clocks in order to make fake tracks very unlikely (Nagamine 2003). As an example, our telescopes (Fig. \ref{TelescopePicture}) have at least three detection matrices and a time resolution of 1 ns (Lesparre et al. 2012a; Marteau et al. 2012).

Fake tracks may however not be the only source of noise affecting the data. In the present paper, we discuss a source of noise produced by upward flux of muons, i.e. muons coming from below the horizontal plane and passing through the telescope. This upward noise perfectly mimics a muon that would have crossed the structure under study since it passes all standard geometrical cuts used to remove fake tracks. Consequently, the only way to remove this noise is to use high-resolution time-of-flight criterion to distinguish particles coming from backward from those actually coming from the volcano.

We identified upward noise in the early data taken on Mount Etna and on La Soufri\`{e}re of Guadeloupe (Lesparre et al. 2012c). In particular, we observed that this noise is present when the telescopes are located on the steep flanks of the volcanoes, with their rear-side turned toward large and deep valleys (e.g. Fig. \ref{AcquisitionSites}). The important role of the atmospheric volume located below and behind the telescope is established through a dedicated experiment performed on La Soufri\`{e}re. In a second part of the paper, we discuss the impact of the measured upward noise on the quality of the density radiographies. For an upward noise to exist in this type of field conditions we must suppose that either muons are produced in the atmosphere volume located below the telescope level or that muons belonging to almost horizontal showers are sufficiently scattered to have an upward trajectory. These explanations deserve dedicated modellings that will be the subject of forthcoming studies.

\section{Field telescopes}

\begin{figure}
\begin{center}
\includegraphics[width=1\linewidth]{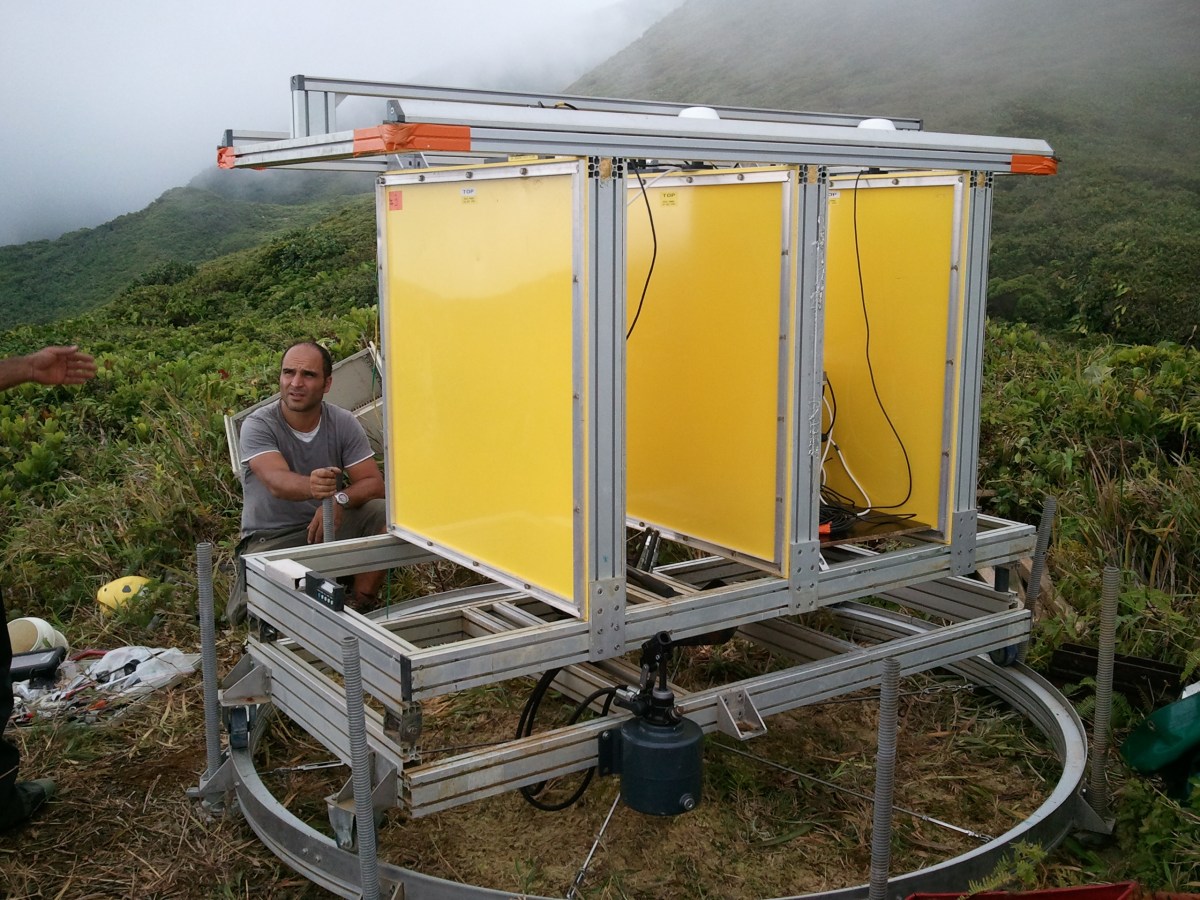}
\includegraphics[width=1\linewidth]{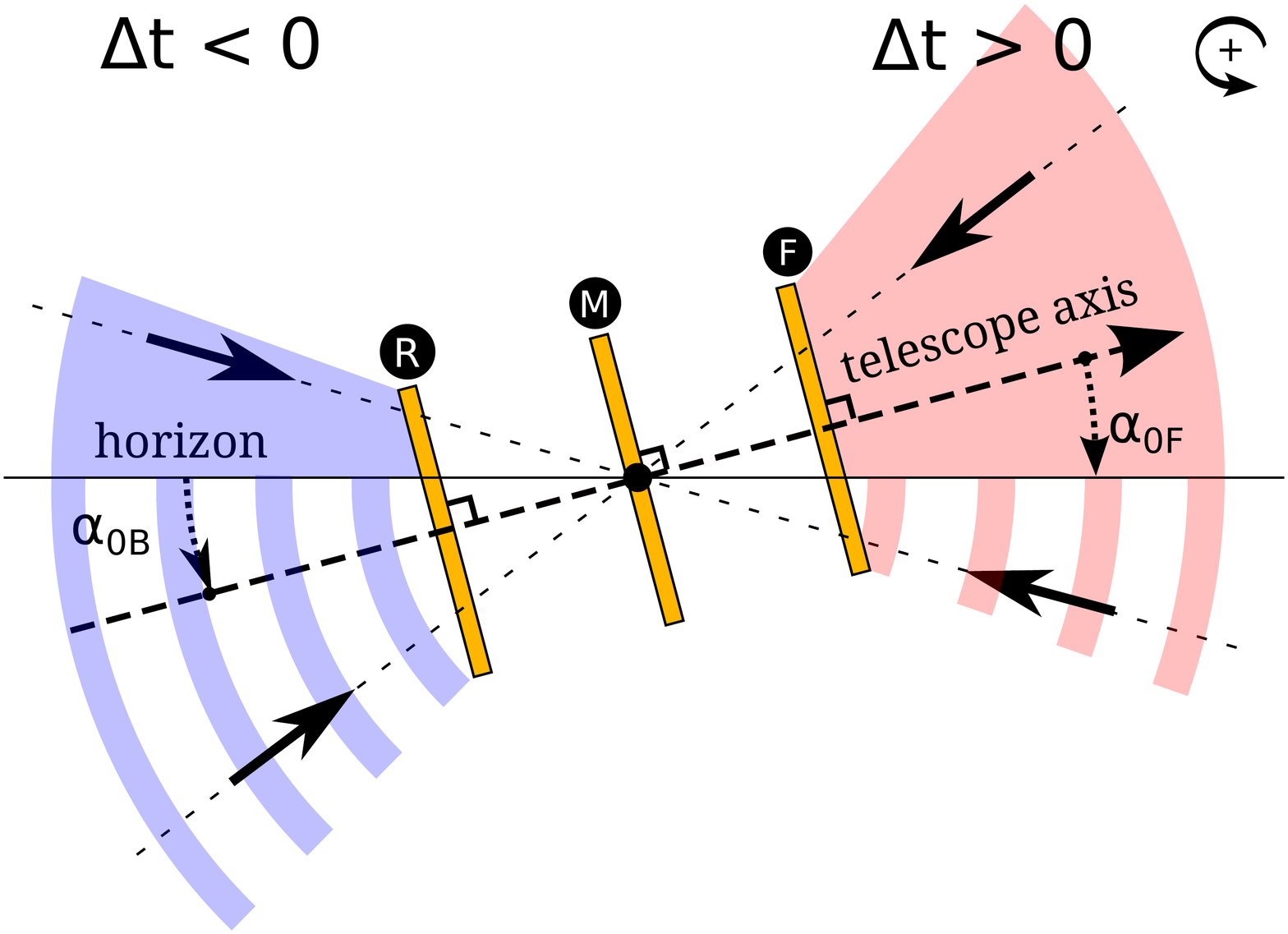}
\vspace{-8mm}
\end{center}
\caption{Top: Picture of a muon telescope in horizontal position. The three detection matrices are in the yellow frames. Bottom: Schematic side view of the matrices (F = front, M = middle, R = rear) showing the different fluxes crossing the telescope. The sign of the time-of-flight $\Delta t = t_{\mathrm{rear}} - t_{\mathrm{front}}$ allows to distinguish a front flux ($\Delta t > 0$) from a rear flux ($\Delta t < 0$). Inclination $\alpha$ of a particle trajectory is measured positive upward and determined from the pixels fired in the three matrices. Inclination is labelled $\alpha_\mathrm{F}$ and $\alpha_\mathrm{B}$ for front flux and back flux respectively.}
\label{TelescopePicture}
\end{figure}

The telescopes used during these experiments are identical and equipped with three matrices labelled $A$, $B$ and $C$ from front to rear and counting $256$ pixels of $5 \times 5 \; \mathrm{cm}^2$ formed by intersecting $N = 16$ horizontal with $N = 16$ vertical scintillator strips (Fig. \ref{TelescopePicture}). A detailed description of these instruments is given by Lesparre et al. 2012a and Marteau et al. (2012). We focus here on the upgrade of the timing system that allows the high-resolution time-of-flight (TOF) analysis performed in the present study.

\subsection{Readout and timing system}

Each matrix is readout by its own independant opto-electronics system. The front-end electronics is auto-triggered and each hit in a matrix is timestamped before being transmitted to the event-building computer embedded in the telescope. The standard event timestamping procedure is based on the latching of the current local counter value with steps of 10ns. This $100 \; \mathrm{MHz}$ clock is generated in a field programmable gate array (FPGA) \textit{via} a phase-locked loop from a precise, stable, common $20 \; \mathrm{MHz}$ clock provided by an external ``master clock'' card. The three matrices are therefore synchronized on the same master clock signal to avoid the inter-clocks drift. The $10 \; \mathrm{ns}$ accuracy obtained is enough for the standard use but too short for the TOF analysis necessary in the present study.

Since the system was not designed initially neither for precise time measurements nor for particles TOF determination, no specific hardware was foreseen. At that point there are two different possibilities: either an increase of the internal clocks (at the cost of a huge power consumption and possible instabilities of the system) or the use of time-to-digital converter (TDC) techniques without extra hardware. TDC modules have been designed and integrated in the existing FPGA of each matrix sensor.

\subsection{Ring oscillator TDC implementation in FPGA}

With a standard counter the measurement of a time interval rely on the number of clock counts recorded. The intrinsic error on the measurement equals the quantization error and the resolution is limited to the clock period. The idea is to use a temporal vernier with two slightly different and controllable frequencies $T_\mathrm{slow}$ and $T_\mathrm{fast}$. The first one is launched on the rising edge of the signal trigger and the second one on the rising edge of the local counter. The determination of the coincidence between the two signals gives access to two values, $N_0$ and $N_1$, representing the number of counts for the slow and the fast clock respectively. Those quasi-similar clocks are obtained thanks to small differences in the routing of the FPGA, allowing timing resolution down to some tens of pico-seconds.

\begin{figure}
\begin{center}
\includegraphics[width=\linewidth]{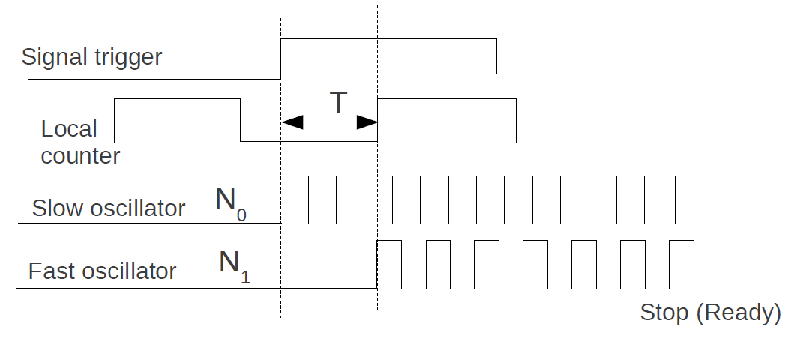}
\end{center}
\caption{Ring oscillator TDC principle.}
\label{fig:tdc} 
\end{figure} 

The timestamp of the signal, with respect to the local counter, is therefore given by,
\begin{equation}
T = (N_0- N_1).T_\mathrm{slow} + N_1.\delta t ,
\end{equation}
where $\delta t = T_\mathrm{slow} - T_\mathrm{fast}$ represents the new resolution of the system. The working principle of this TDC technique and the schematics of its implementation are displayed in Fig.~\ref{fig:tdc}. A phase detector is used to determine the coincidence between the two oscillators. With such a system the timestamp accuracy was improved from $10 \; \mathrm{ns}$ down to $250 \; \mathrm{ps}$. Calibration procedures allow measuring for each sensor the parameters $T_\mathrm{slow}$ and $\delta t$. The values of those parameters may diverge from one sensor to the other due to the small dispersion in the FPGA internal layout. After this upgrade, the overall timing resolution of the apparatus is driven by the responses of the scintillator, the optical fibres and the photomultiplier. In average a global resolution of $1-2 \; \mathrm{ns}$ has been measured with an oscilloscope.

\section{Field experiments}

This section presents three experiments whose data are used to study samples of upward noise. Two experiments were performed on La Soufri\`{e}re of Guadeloupe volcano (Lesparre \textit{et al.} 2012) and a third one on Mount Etna (Table \ref{TableSites}). Auxiliary data in the Rennes University (Britanny, France) laboratory are also used for calibration. Before presenting them into detail, let's precise the angle notation adopted in the present study. The telescope main axis is defined as the vector perpendicular to the three detection matrices planes and oriented from rear to front (Fig. \ref{TelescopePicture}). The orientation of the telescope is defined by the azimuth, $\beta_0$, and the zenith angle $\gamma_0$ of its main axis. Azimuth is measured positive eastward and is the angle from North to the horizontal projection of the telescope axis, for instance $\beta_0 = 90^\circ$ when the telescope looks eastward. Zenith angle is always positive and measured from the upward vertical axis to the telescope axis, i.e. $\gamma_0 = 90^\circ$ when the telescope axis is horizontal as shown in the picture of Fig. \ref{TelescopePicture}.

Fig. \ref{TelescopePicture} represents the different fluxes distinguished in the present study. The forward flux $\phi_\mathrm{\textsc{f}}$ enters in the telescope through the front matrix and escapes through the rear matrix. Conversely, the backward flux $\phi_\mathrm{\textsc{b}}$ enters through the rear matrix and escapes through the front matrix. We also distinguish upward and downward fluxes respectively represented by $\phi_\mathrm{\textsc{f,b}}^u$ and $\phi_\mathrm{\textsc{f,b}}^d$.

Because of the particular role played by the horizontal plane in the analysis, we also refer to slope angles, $\alpha$, defined relative to the horizontal plane within the telescope frame and measured positive upward. Negative slopes are consequently assigned to telescope lines of sight looking below the horizontal plane as shown in Fig. \ref{AcquisitionSites}. When necessary, we distinguish slopes $\alpha_\mathrm{\textsc{f}}$ in the forward direction and slopes $\alpha_\mathrm{\textsc{b}}$ in the backward direction with the following equivalences (Fig. \ref{TelescopePicture}),
\begin{equation}
\alpha_\mathrm{\textsc{f}} = \gamma - \frac{\pi}{2} = -\alpha_\mathrm{\textsc{b}}.
\end{equation}
With these notations, the different fluxes are such that,
\begin{subequations}
\begin{align}
\phi_\mathrm{\textsc{f}}^u = \phi_\mathrm{\textsc{f}}(\alpha_\mathrm{\textsc{f}} > 0), \\
\phi_\mathrm{\textsc{f}}^d = \phi_\mathrm{\textsc{f}}(\alpha_\mathrm{\textsc{f}} < 0), \\
\phi_\mathrm{\textsc{b}}^u = \phi_\mathrm{\textsc{b}}(\alpha_\mathrm{\textsc{b}} > 0), \\
\phi_\mathrm{\textsc{b}}^d = \phi_\mathrm{\textsc{b}}(\alpha_\mathrm{\textsc{b}} < 0).
\end{align}
\end{subequations}

\begin{table*}
\caption{Site characteristics and main acquisition parameters. Note that for Etna site: apertures are for ranges of $15 \; \mathrm{km}$ ($\dag$) and $30 \; \mathrm{km}$ ($\ddag$). Also for the high definition data sets \textsc{smtomo} and \textsc{etomo} we specify the characteristics of each sub-dataset when different.}
\label{TableSites}
\begin{center}
\begin{tabular}{lcccc}
\hline
  & Rennes Lab & Roche Fendue & Savane \`a Mulets & Etna\\
Data set & \textsc{lbcalib} & \textsc{rfcalib} & \textsc{smtomo} & \textsc{etomo} \\
\hline
Altitude a.s.l.                	& 56 m		& 1268 m        & 1189 m       & 3095 m \\
$X_{UTM}$ WGS84                	& (30)601373 m	& (20)643347 m  & (20)642599 m & (33)499007 m\\
$Y_{UTM}$ WGS84                	& 5330300 m	& 1774036 m     & 1773852 m    & 4178852 m \\
Zenith angle $\gamma_0$        	& $90^\circ$	& $90^\circ$    & $80^\circ$ - $85^\circ$ - $85^\circ$   & $85^\circ$ - $80^\circ$ \\
Forward azimuth $\beta_0$      	& $68.6^\circ$	& $215.5^\circ$ & $32^\circ$ - $44^\circ$ - $56^\circ$   & $110^\circ$ \\
Open space aperture            	& $0^\circ$	& $-7.4^\circ$  & $-7.4^\circ$  & $-8.0^\circ \; ^\dag$ \\
\hspace{3 mm} below horizontal 	&		& & & $-3.9^\circ \; ^\ddag$ \\
Number of matrices             	& 3		& 3 & 3 & 3 \\
Matrix distance $D$            	& 85 cm		& 85 cm & 60 cm & 85 cm \\
Axial acceptance               	& $5.8 \; \mathrm{cm}^2\mathrm{sr}$ & $5.8 \; \mathrm{cm}^2\mathrm{sr}$ & 11.5 $\; \mathrm{cm}^2\mathrm{sr}$ & $5.8 \; \mathrm{cm}^2\mathrm{sr}$ \\
Angular resolution             	& $1.6^\circ$   & $1.6^\circ$ & $2.5^\circ$ & $1.6^\circ$ \\
Acquisition time               	& 7 days	& 4 days & 22 - 27 - 17 days & 7 - 4 days \\
\hline  
\end{tabular}
\end{center}
\end{table*}

\subsection{The Rennes\,1 University laboratory site}

The Rennes\,1 University site is located on the third floor of building 15 of the Beaulieu campus, at and altitude of $56 \; \mathrm{m}$ about $12 \; \mathrm{m}$ above the ground (Table \ref{TableSites}). The topography may reasonably be considered flat in a wide region of tens of kilometres around the telescope with the nearest noticeable topography highs ($\approx 250 \; \mathrm{m}$) located $160 \; \mathrm{km}$ and $40 \; \mathrm{km}$ away in the forward and backward direction respectively.

The data set acquired at this location is referred to as \textsc{lbcalib} and used to calibrate and test the telescope previously installed on Mount Etna. The telescope is oriented horizontally (i.e. $\gamma_0 = 90^\circ$). The telescope does not face any significant obstruction excepted some concrete and plaster walls representing half a metre of concrete, and there is no open space aperture below the horizontal.

\subsection{La Soufri\`{e}re Roche Fendue site}

The Roche Fendue site is located on the Eastern side of La Soufri\`{e}re lava dome at an altitude of $1268 \; \mathrm{m}$ (Table \ref{TableSites}).

The corresponding data set, called \textsc{rfcalib}, is dedicated to upward flux detection with the telescope oriented horizontally with its front face oriented in the South-West direction along the axis of a small valley with a gentle slope leading to a wide and deep valley located southward of La Soufri\`{e}re (Fig. \ref{AcquisitionSites}). The front side of the telescope sees an open space free of rock obstruction down to $-6^\circ$ (Top part of Fig. \ref{AcquisitionSites}). Beyond $-6^\circ$ the lines of sight encounter the Caribbean Mounts and the rock obstruction progressively increases up to $1.5 \; \mathrm{km}$ at $-7.4^\circ$. The backward landscape begins with a small horizontal plateau formed by volcanic deposits at the Col de l'\'{E}chelle located East of La Soufri\`{e}re. This plateau produces a small rock obstruction of about $0.13 \; \mathrm{km}$ that remains constant in $[-6.8^\circ; 0^\circ]$ and progressively decreases to zero at $7^\circ$ above the horizontal (Top part of Fig. \ref{AcquisitionSites}). Consequently, the downward backward flux will also be slightly obstructed for this data set.

\subsection{La Soufri\`{e}re Savane \`a Mulets site}

\begin{figure*}
\begin{center}
\includegraphics[width=0.33\linewidth]{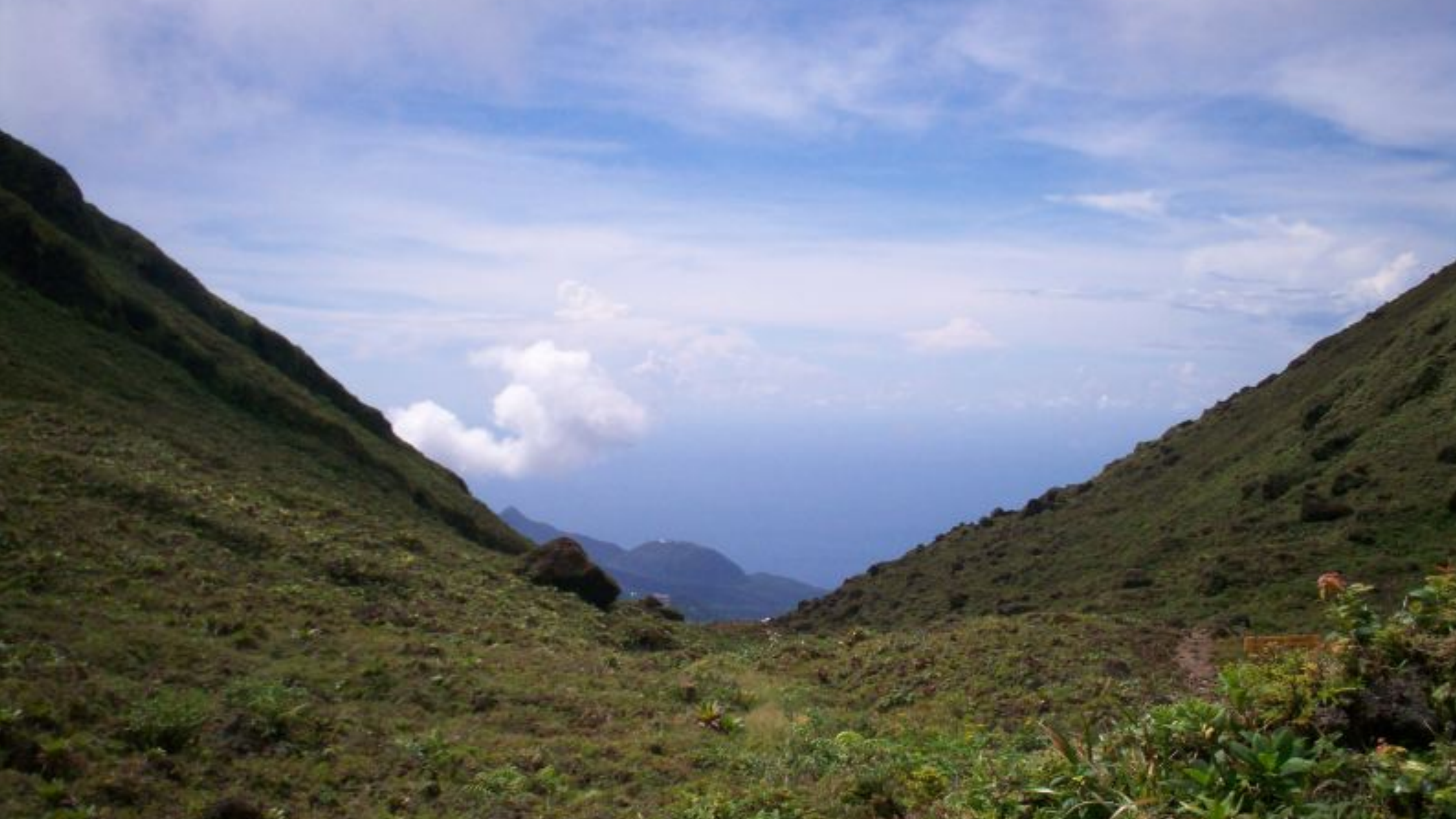}
\includegraphics[width=0.33\linewidth]{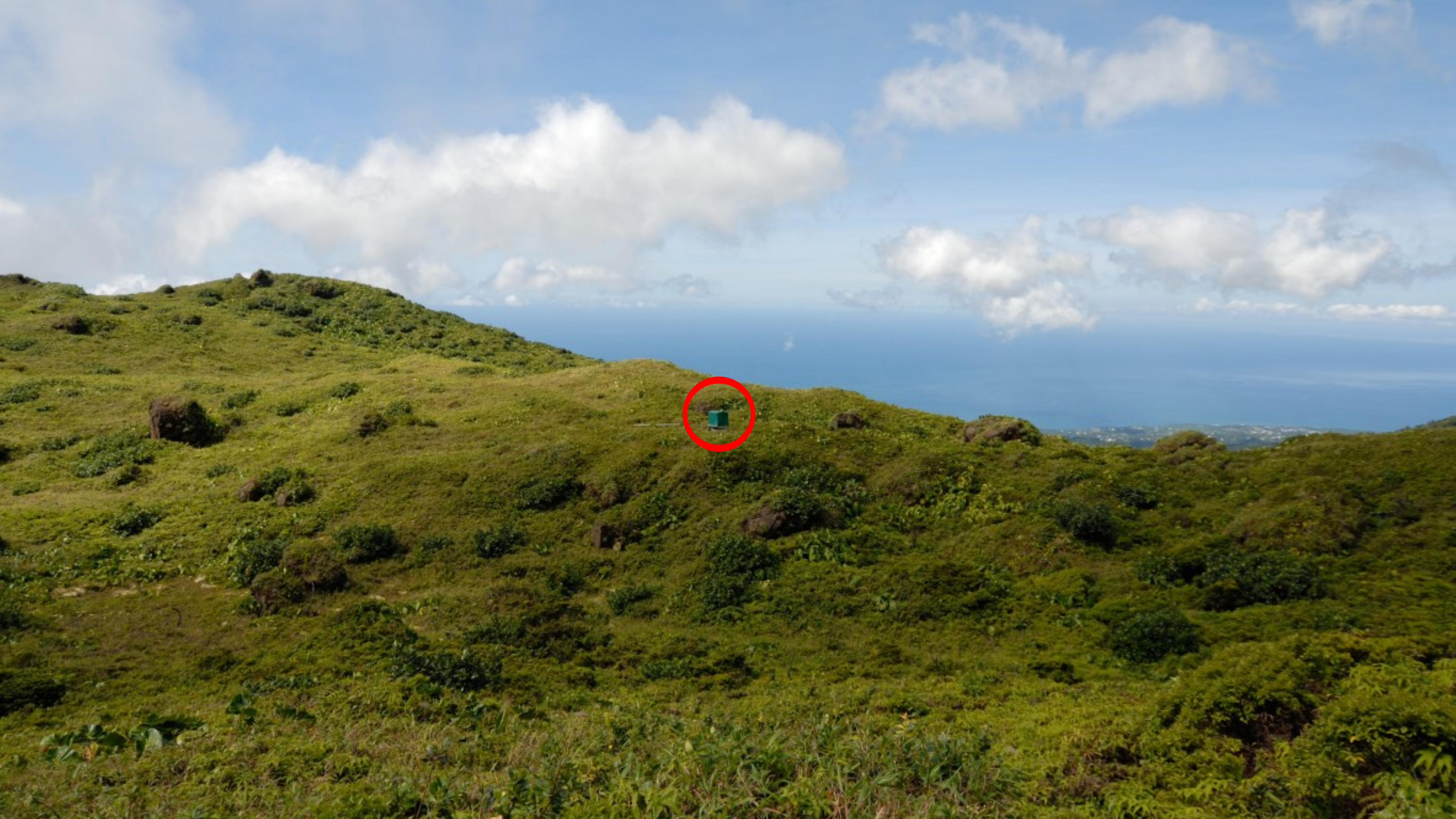}
\includegraphics[width=0.33\linewidth]{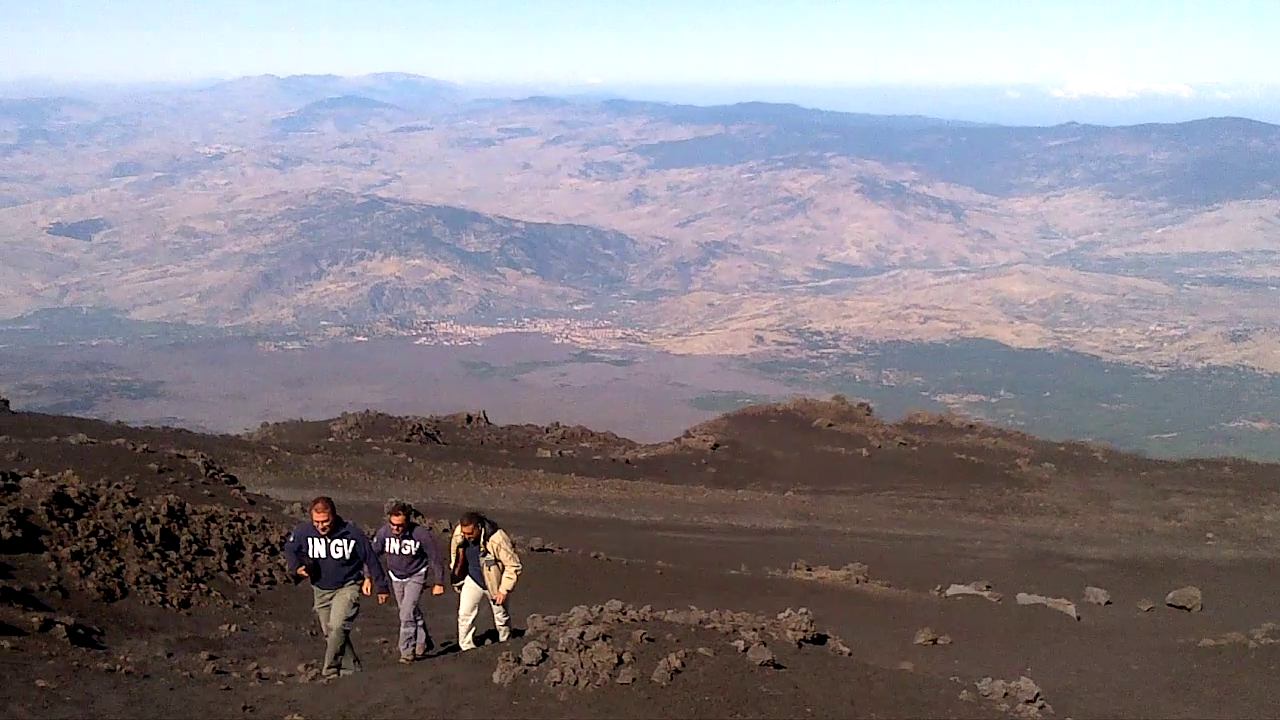}
\includegraphics[width=1\linewidth]{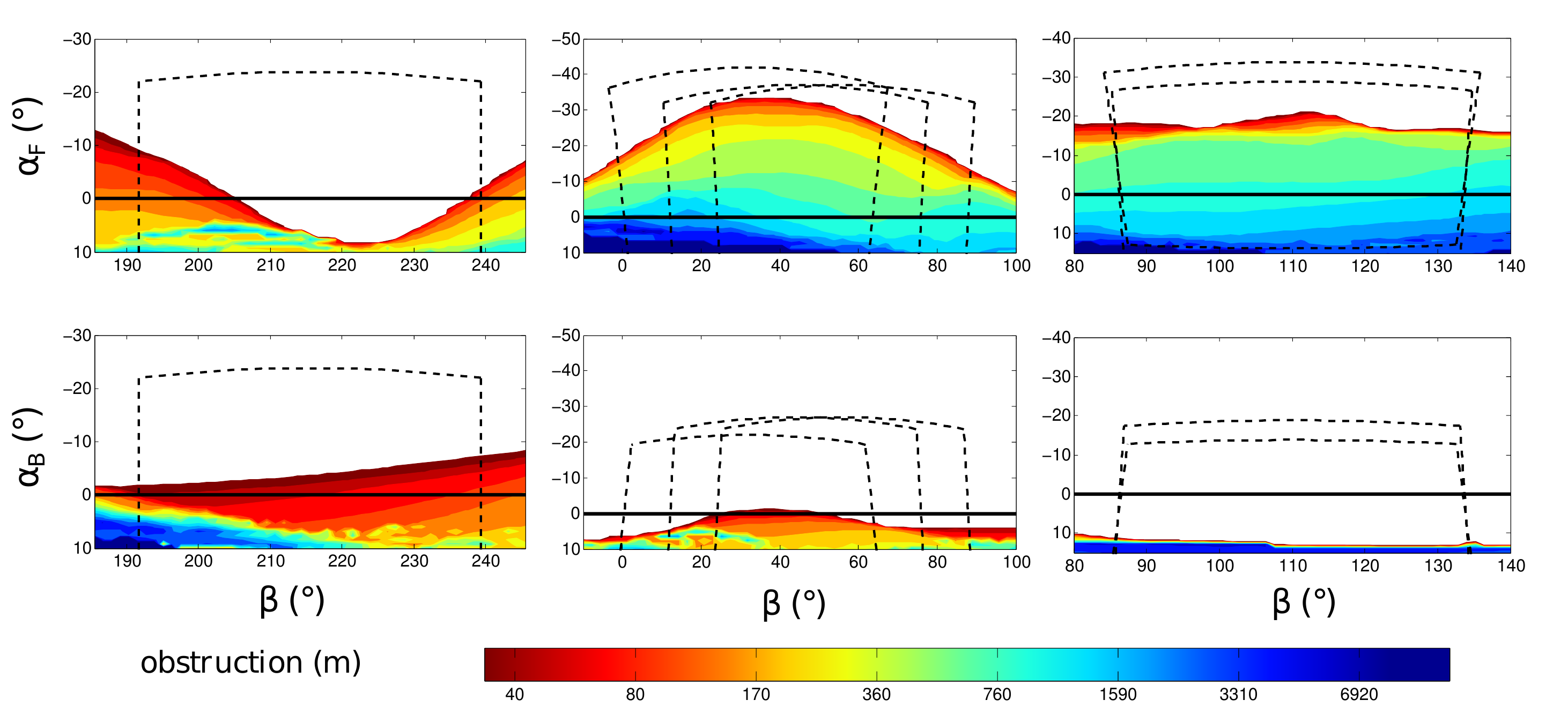}
\includegraphics[width=1\linewidth]{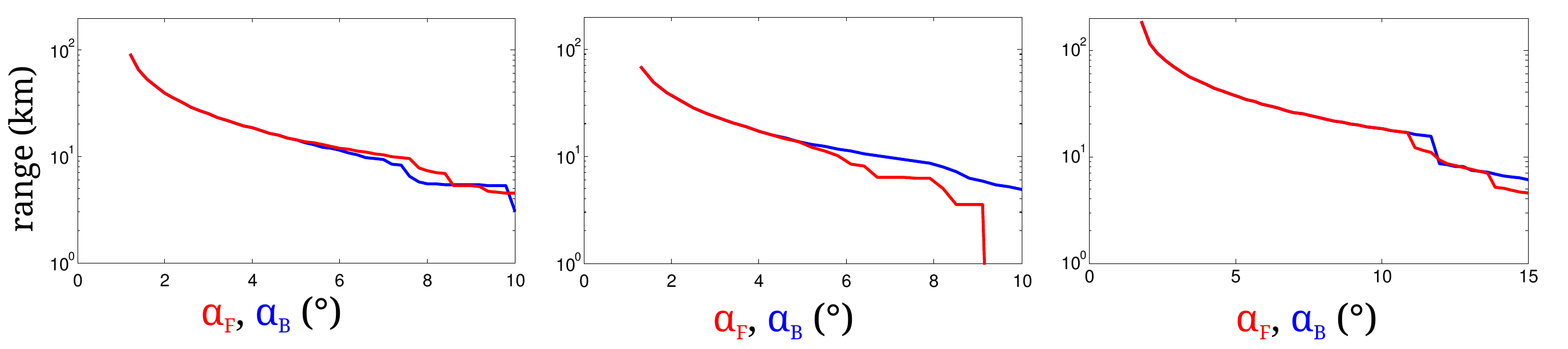}
\vspace{-8mm}
\end{center}
\caption{
TOP: Left -- Picture taken southward (i.e. forward direction) from the Roche Fendue site. The mount located 8 km away in the valley axis if the Hou\"elmont that hosts the Volcano Observatory at an altitude of approximately 400 m. Middle -- Wide view of the Savane \`a Mulets site showing the telescope (circle) with low-altitude topography and the Caribbean sea in the background. Right -- View taken in the backward direction from the Etna North-East crater site showing the open space seen by the back side of the telescope. MIDDLE: Rock obstruction (m) facing the telescope in the different acquisition configurations. \textsc{rfcalib} on the left, \textsc{smtomo} on the middle and \textsc{etomo} on the right. The rock obstruction is defined as the total rock thickness encountered along a given line of sight. The telescope acquisition zone is delimited by a dotted black line. BOTTOM : Range (km) for the different acquisitions along the central azimuth of the telescope (it does not fluctuate importantly with the azimuth). The range is the distance from the telescope where the line of sight definitely enters into the Earth.}
\label{AcquisitionSites}
\end{figure*}

The Savane \`{a} Mulets site is located at the edge of a narrow plateau located beneath the western side of the lava dome at an altitude of $1189 \; \mathrm{m}$ (Fig. \ref{AcquisitionSites}).

The data set, called \textsc{smtomo}, was acquired during a high-resolution tomography experiment of the lava dome and merges three data acquisitions made at the same location with different azimuth angles and slightly different zenith angles (Table \ref{TableSites}). The topography profile taken along the axial line of sight of the telescope (bottom of Fig. \ref{AcquisitionSites}) shows that the front side of the telescope sees a landscape obstructed by La Soufri\`ere that produces an obstruction that varies from $1.75 \; \mathrm{km}$ at $0^\circ$ to $3 \; \mathrm{km}$ at $-4.1^\circ$. The backward side of the profile is directed toward a wide open space with an obstruction that varies from $0.1 \; \mathrm{km}$ at $0^\circ$ to $0.3 \; \mathrm{km}$ at $-6^\circ$. Beyond this slope, the Caribbean Mounts produce a stronger obstruction that rapidly increases to $1.8 \; \mathrm{km}$ at $-7.5^\circ$.

\subsection{The Etna North-East crater site}

This site is located at an altitude of $3095 \; \mathrm{m}$, about three times higher than La Soufri\`{e}re sites. The back side of the telescope is directed toward a deep open space down to $-8.0^\circ$ below the horizontal for distance up to $15 \; \mathrm{km}$. For a range of $30 \; \mathrm{km}$, the slope of the open space is reduced to $-3.9^\circ$ (Table \ref{TableSites}). The data set, called \textsc{etomo}, merges the data for two acquisitions performed with a constant azimuth angle and slightly different zenith angles.

\section{Analysis method}

\subsection{Statistical detection of an upward flux}

We now describe the method used to detect the existence of an upward flux in the data sets. The analysis relies on the use of geometrical information provided by the pixelated matrices of the telescope and time information furnished by the high-speed electronics.

The geometrical information is the muon trajectory determined by the pair of pixels $(a_{i,j}, b_{k,l})$ fired by the particle. Pixel $a_{i,j}$ belongs to the front matrix $A$ and $c_{k,l}$ belongs to the rear matrix $C$. Indexes $i,k,j,l$ vary from $1$ to $N = 16$, and the combination of all possible pairs of pixels $(a_{i,j}, c_{k,l})$ defines a set of  $(2 N - 1)^2 = 961$ discrete directions of sight $\mathbf{r}_{m,n}$ with $m = i-k$ and $n = j-l$ (Lesparre \textit{et al.} 2012a; Marteau \textit{et al.} 2012). The angular range spanned by the $961$ directions may be controlled by adjusting the distance $2 \times D$ between the front and rear matrices and, for the present experiments, this distance is set to either $120$ or $170 \, \mathrm{cm}$ depending on the site (Table \ref{TableSites}).

One particle is considered detected when one pixel is fired on each matrix in a $20~\mathrm{ns}$ time interval. The telescope returns a flux in $\mathrm{s}^{-1}$ that is corrected from the instrument acceptance and efficiency to recover the absolute flux ${\phi}$ in $\mathrm{s}^{-1}.\mathrm{sr}^{-1}.\mathrm{cm}^{-2}$ (Lesparre et al. 2012b).

For one direction of detection $\mathbf{r}_{m,n}$, the total flux recorded by the telescope reads
\begin{equation}
{\phi}_{m,n} = {\phi}_{m,n}^{d} + {\phi}_{m,n}^{u},
\label{eqn_total1}
\end{equation}
where ${\phi}_{m,n}^{d}$ is the downward flux coming from over the horizon and ${\phi}_{m,n}^{u}$  is the upward flux entering the opposite side of the telescope and coming from below the horizon. In order to display the results in a common geographical frame independent of both measurement site and telescope configuration, we replace $\mathbf{r}_{m,n}$ with slopes ${\alpha}$ and azimuth angles ${\beta}$. Using this convention, eq. (\ref{eqn_total1}) gives two possibilities,
\begin{equation}
\phi_{m,n} =
\begin{array}{l}
\phi_{\alpha_\mathrm{F},\beta}^d + \phi_{-\alpha_\mathrm{F},\pi+\beta}^u \\
\; \\
\phi_{\alpha_\mathrm{F},\beta}^u + \phi_{-\alpha_\mathrm{F},\pi+\beta}^d.
\end{array}
\label{eqn_total2}
\end{equation}
depending on whether the forward line of sight has a positive or a negative slope $\alpha_\mathrm{\textsc{f}}$.

A way to use eq. (\ref{eqn_total2}) to distinguish ${\phi}^{d}$ from ${\phi}^{u}$ is to record two data sets for $(\alpha,\beta)$ and $(\alpha,\pi+\beta)$ with an open-sky field configuration such that an upward flux is expected to only come from direction $(\pi-\alpha,\pi+\beta)$. From eq. (\ref{eqn_total2}), we have:
\begin{align}
{\phi}_{\alpha_F,\beta} & = {\phi}_{\alpha_F,\beta}^{d} + {\phi}_{-\alpha_F,\pi+\beta}^{u} \label{eqn_diffFlux1a} \\
{\phi}_{\alpha_F,\pi+\beta} & = {\phi}_{\alpha_F,\pi+\beta}^{d}.
\label{eqn_diffFlux1b}
\end{align}
Assuming an azimuthal symmetry of the open-sky flux, i.e. ${\phi}_{\alpha_F,\beta}^{d} = {\phi}_{\alpha_F,\pi+\beta}^{d}$, eq. (\ref{eqn_diffFlux1a}) and (\ref{eqn_diffFlux1b}) give,
\begin{equation}
{\phi}_{\alpha_F,\beta} - {\phi}_{\alpha_F,\pi+\beta} = {\phi}_{-\alpha_F,\pi+\beta}^{u}.
\label{eqn_diffFlux2}
\end{equation}
However, even for the particularly favourable field situation of \textsc{rfcalib} (see Table \ref{TableSites}), we realized that eq. (\ref{eqn_diffFlux2}) is of limited practical usage for two reasons. Indeed, beside the fact that field configurations with no upward flux in one direction are quite rare, the muon flux is sensitive to the geomagnetic field (Hansen et al. 2005) and ${\phi}_{\alpha,\beta}$ may slightly differ from ${\phi}_{\alpha,\pi + \beta}$. This discrepancy is more important at high zenith angles because the muons spend more time interacting with the geomagnetic field.

We therefore use information brought by the telescope muon time-of-flight defined as ${\Delta t} = {t}_\mathrm{rear} - {t}_\mathrm{front}$ where ${t}_\mathrm{front}$ and ${t}_\mathrm{rear}$ are the dates when the particle cross the front and rear matrices respectively (Fig. \ref{TelescopePicture}). Particles coming from the forward direction have a ${\Delta t} > 0$, backward particles have $\Delta t < 0$, and the TOF, $\Delta t^{u}$, of an upward particle is the opposite of the TOF, $\Delta t^{d}$, of the corresponding downward particle. For a direction $\mathbf{r}_{m,n}$, the theoretical unsigned TOF is given by,
\begin{equation}
{\Delta t}_\mathrm{theo}{(m,n)} = \frac{\sqrt{4 \times D^{2} + (m^2 + n^2) \times \delta^2}}{c},
\label{eqn_tEpec}
\end{equation}
where $\delta = 5 \; \mathrm{cm}$ is the pixel size and $c$ is the light speed in air.

\begin{figure}
\begin{center}
\includegraphics[width=\linewidth]{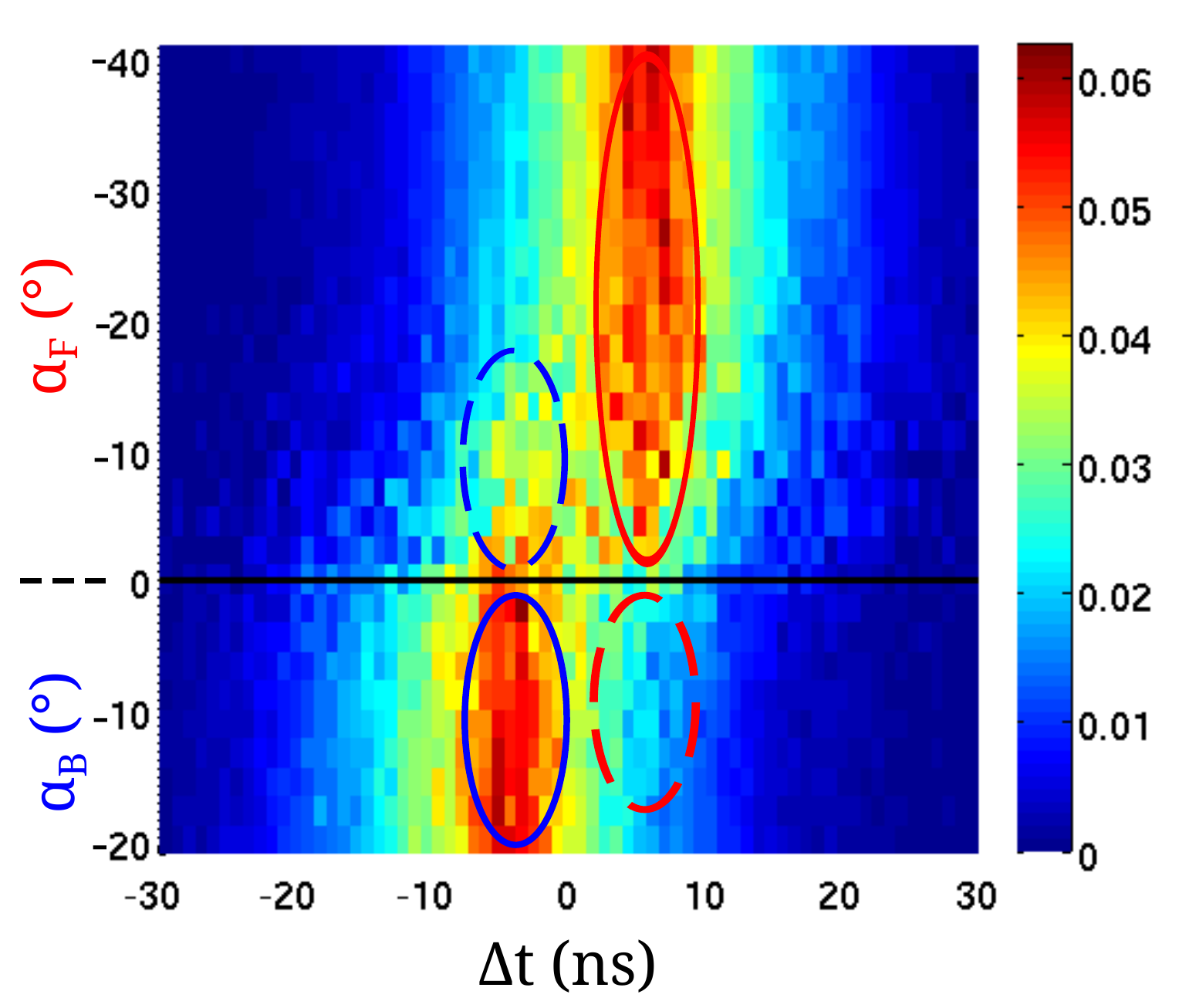}
\vspace{-8mm}
\end{center}
\caption{TOF distribution for the \textsc{smtomo} data set shown as normalized histograms as a function of zenith angle. The horizon is represented by the dashed line. The blue and red solid ellipses respectively show the backward ($\alpha_B < 0$ and $\Delta t < 0$) and forward ($\alpha_F < 0$ and $\Delta t > 0$) events corresponding to the downward fluxes. The dashed ellipses show events corresponding to upward-going muons from forward (red ellipse, $\alpha_B < 0$ and $\Delta t > 0$) and backward (blue ellipse, $\alpha_F < 0$ and $\Delta t < 0$).}
\label{telescopeFlyingTimeHistosVsZenithAngle}
\end{figure}

Despite the high-speed electronics used in the telescopes, time measurements are altered by several uncertainties of different types that prevent a deterministic particle-by-particle TOF determination. These uncertainties come from latencies of photon generation in the scintillator (e.g. Bross \textit{et al.} 1993; Pla-Dalmau \textit{et al.} 2001), the chaotic capture of photons by the optical shifting fibres and their propagation along the fibres down to the photomultipliers (e.g. Kudenko \textit{et al.} 2001). These phenomena produce a dispersion of the ${\Delta t}$'s in a range of $\approx 10 \; \mathrm{ns}$ centred on ${\Delta t}_\mathrm{theo}$ (Fig. \ref{telescopeFlyingTimeHistosVsZenithAngle}).

The TOF probability density distribution, $\mathcal{P}(\Delta t)$, may be decomposed as,
\begin{equation}
\mathcal{P}(\Delta t) = \mathcal{P}^d(\Delta t) + \mathcal{P}^u(\Delta t),
\label{eqn_probMixt1}
\end{equation}
where $\mathcal{P}^d$ and $\mathcal{P}^u$ respectively correspond to the distributions of the downward and upward TOF's. These distributions have the same normalized shape $\mathcal{F}(\mathrm{mean},\mathrm{std})$, as they share the same causes of uncertainties, but they have different amplitude and opposite mean. Using this property, eq. (\ref{eqn_probMixt1}) rewrites as,
\begin{equation}
\mathcal{P}(\Delta t) = r^{d} \, \mathcal{F}({\Delta t}_\mathrm{theo},\sigma) + (1-r^{d}) \, \mathcal{F}({-\Delta t}_\mathrm{theo},\sigma)
\label{eqn_probMixt2}
\end{equation}
where $0 \le r^{d} \le 1$ is the downward flux ratio equals to $0$ for a pure upward flux and to $1$ for a pure downward flux. From eq. (\ref{eqn_probMixt2}), the average of the measured TOF's reads,
\begin{equation}
\langle {\Delta t} \rangle = {\Delta t}_\mathrm{theo}\times (r^{d} - (1-r^{d})),
\label{eqn_ratio2}
\end{equation}
and
\begin{equation}
r^{d} = \frac{\langle {\Delta t} \rangle + {\Delta t}_\mathrm{theo}}{2 \, {\Delta t}_\mathrm{theo}}.
\label{eqn_ratio3}
\end{equation}
In the remaining we shall distinguish the forward ratio, $r_\mathrm{\textsc{f}}^d = \phi_\mathrm{\textsc{f}}^d / \phi_\mathrm{\textsc{b}}^u$, and the backward ratio, $r_\mathrm{\textsc{b}}^d = \phi_\mathrm{\textsc{b}}^d / \phi_\mathrm{\textsc{f}}^u$.

Now that we introduced all the variables, we sum up the procedure to recover the muon upward flux:
\begin{itemize}
\item
For each direction of observation $\mathbf{r}_{m,n}$ the measured TOF ${\Delta t}$ are bootstrapped to get ${\langle {\Delta t} \rangle}$ (eq. \ref{eqn_ratio2}) and its uncertainty.
\item
The theoretical TOF is computed with eq. (\ref{eqn_tEpec}).
\item
The downward flux ratio $r^d$ and its uncertainty are derived with eq. (\ref{eqn_ratio3}). Eventually, $r^d$ may be averaged over azimuthal angle to be written as a function of zenith angle only. This simplification is valid only when obstruction is the same at all azimuths considered.
\item
Finally, we recover,
\begin{subequations} \label{phiDU}
\begin{align}
\phi^{d} & = r^{d} \times \phi, 
\label{eqn_ratio4} \\
\phi^{u} & = (1-r^{d}) \times \phi.
\end{align}
\end{subequations}
\end{itemize}

One needs to make a clear distinction between the ratio $r^{d}$ and the fluxes ${\phi}$ of eq. (\ref{phiDU}). These quantities are computed with different information and have different uncertainties. The fluxes $\phi^{d}$ and $\phi^{u}$ cumulate the uncertainties brought by both $r^d$ and $\phi$, and $r^{d}$ is a better parameter to prove the existence of an upward flux. However $r^{d}$ is not an intuitive parameter to characterize the upward flux as obstruction on $\phi^{d}$ and/or $\phi^{u}$ path will enhance or diminish it.

\section{Data analysis} \label{DataAnalysis}

\subsection{Calibration data sets}

The calibration data sets analysed in the present study are acquired by setting the telescope in horizontal position in order to have identical parameters -- same acceptance and observation axis -- in both the forward and backward directions. When in horizontal position, the telescope lines of sight $\mathbf{r}_{m,0}$ encompass a small range of slopes, $(-\delta\alpha, +\delta\alpha)$, where all types of fluxes ($\phi_\mathrm{\textsc{f}}^u$, $\phi_\mathrm{\textsc{b}}^u$, $\phi_\mathrm{\textsc{f}}^d$ and $\phi_\mathrm{\textsc{b}}^d$) are indifferently collected and indistinguishable.

The \textsc{lbcalib} data set corresponds to a totally symmetrical configuration with no obstruction for $\alpha > 0$ in both the forward and the backward directions. Since the telescope is surrounded by a very wide flat area, all upward trajectories corresponding to negative slopes directly enter the Earth and have a range limited to few tens of meters. The \textsc{rfcalib} data set corresponds to a non-symmetrical situation with no obstruction up to $\alpha_\mathrm{\textsc{f}} =  6^\circ$ in the forward direction and a small constant obstruction of $0.2 \; \mathrm{km}$ down to to $\alpha_\mathrm{\textsc{b}} = 6.8^\circ$ in the backward direction (Fig. \ref{AcquisitionSites}).

\begin{figure*}
\begin{center}
\includegraphics[width=1\linewidth]{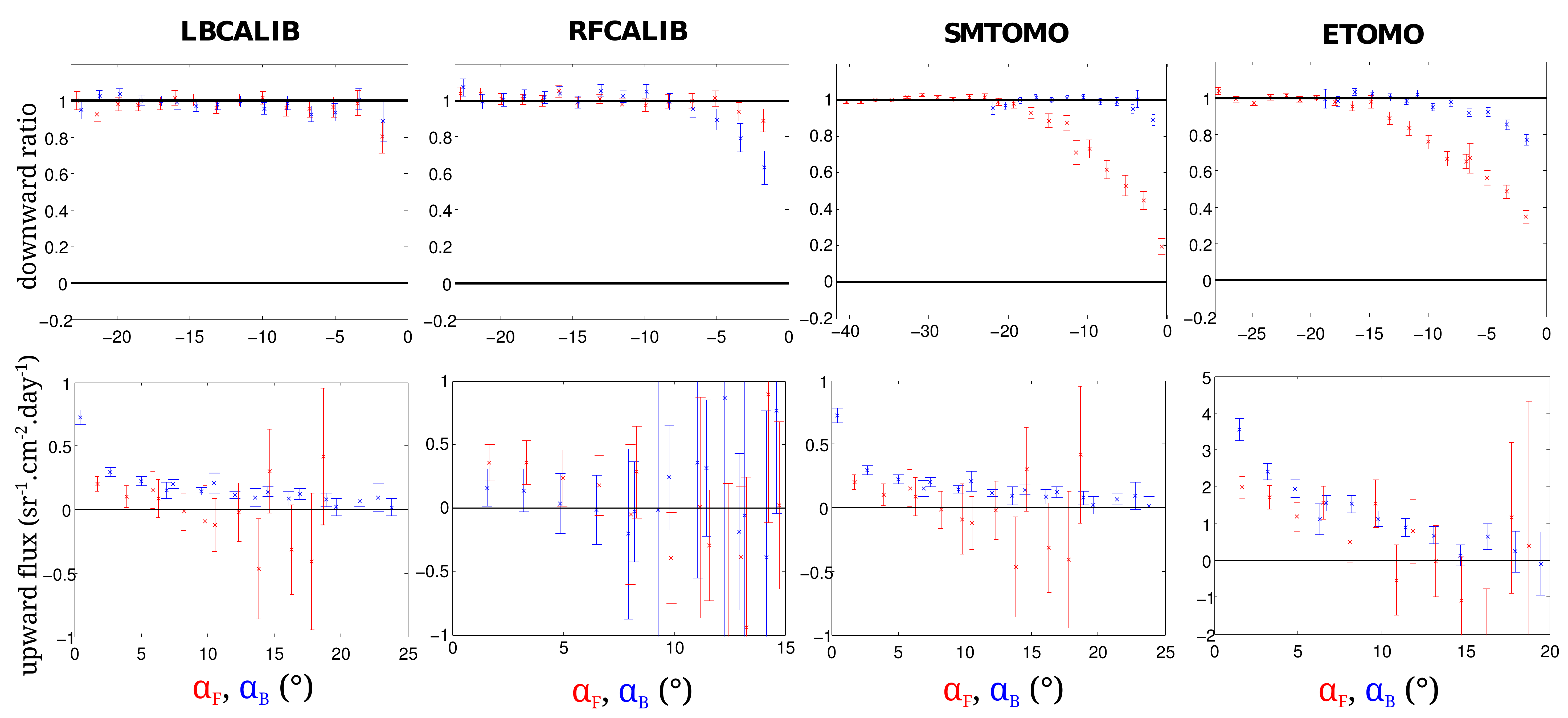}
\vspace{-8mm}
\end{center}
\caption{
Top: downward ratio $r_\mathrm{\textsc{f}}^d = \phi_\mathrm{\textsc{f}}^d / \phi_\mathrm{\textsc{b}}^u$ (red) and $r_\mathrm{\textsc{b}}^d = \phi_\mathrm{\textsc{b}}^d / \phi_\mathrm{\textsc{f}}^u$ (blue). Bottom: upward fluxes $\phi_\mathrm{\textsc{f}}^u$ (red) and $\phi_\mathrm{\textsc{b}}^u$ (blue).
}
\label{FluxRatio}
\end{figure*}

Both the downward ratio $r^d$ and the upward flux $\phi^u$ computed for the \textsc{lbcalib} and \textsc{rfcalib} data sets are shown in Fig. \ref{FluxRatio}. For \textsc{lbcalib} and as expected, $r^d \approx 1$ for all slopes excepted in a narrow range of $ \alpha_\mathrm{\textsc{f,b}} > -3^\circ$ above the horizon where $r^d \lessapprox 1$ on both sides. This corresponds to a tiny upward flux of $0.3~\mathrm{sr}^{-1}.\mathrm{cm}^{-2}.\mathrm{day}^{-1}$.

The $r^d$ values obtained for the \textsc{rfcalib} data set are significantly less than $1$ for $\alpha_\mathrm{\textsc{f}} < -5^{\circ}$ for the forward flux and for $\alpha_\mathrm{\textsc{b}} < -10^{\circ}$ for the backward flux. If the backward $r^d$ is significantly lower than the forward one, it is partially compensated on $\phi^u$ because the forward flux is stronger than the backward flux. Forward $\phi^u$ takes values very close to \textsc{lbcalib} on the first $4^{\circ}$ below the horizon. It then seems to be stagnating at $0.2~\mathrm{sr}^{-1}.\mathrm{cm}^{-2}.\mathrm{day}^{-1}$ until $10^{\circ}$ below the horizon where the noise forbids any reading. Forward $\phi^u$ goes from $0.2~\mathrm{sr}^{-1}.\mathrm{cm}^{-2}.\mathrm{day}^{-1}$ to $0$ on the first $4^{\circ}$ below the horizon. The difference between the forward and the backward $\phi^u$ may be explained by the different obstruction patterns they are facing.

\subsection{Tomography data sets}

During the tomography measurements, the telescope is slightly inclined to have its main axis oriented toward the part of the volcano to be imaged with the largest acceptance. Consequently, the range of slopes spanned for the forward and the backward direction are different, and the forward flux $\phi_\mathrm{\textsc{f}}^d$ is strongly attenuated by the volcano, leading to an enhanced lowering of the downward ratio. Tomography data sets benefit from a good signal-to-noise ratio due to their longer acquisition time (Table \ref{TableSites}).

The azimuthally-averaged ratio $r^d$ computed with the middle \textsc{smtomo} data subset (i.e. $\gamma_0 = 85^\circ$ and $\beta_0 = 44^\circ$) is shown on Fig. \ref{FluxRatio}. The forward obstruction for this data subset goes from a few tens of meters at the upper edge of the volcano to $\approx\,1 \; \mathrm{km}$ for horizontal lines of sight (Fig. \ref{AcquisitionSites}). Below the horizontal plane, obstruction quickly reaches $10 \; \mathrm{km}$. The backward obstruction is null above the horizon and gradually increases up to a few hundred meters at $\alpha_\mathrm{\textsc{b}} = 10^{\circ}$. The discrepancy between the forward and backward obstructions observed for the \textsc{smtomo} data is retrieved in the  corresponding $r^d$ curves (third top plot of Fig. \ref{FluxRatio}). The forward ratio $r_\mathrm{\textsc{f}}^d = \phi_\mathrm{\textsc{f}}^d / \phi_\mathrm{\textsc{b}}^u$ (red curve) takes low values $0.2 < r_\mathrm{\textsc{f}}^d < 0.4$ for $0 > \alpha_\mathrm{\textsc{f}} > -5^{\circ}$ and linearly grows to $0.9$ at $\alpha_\mathrm{\textsc{f}} \approx~-15^{\circ}$. The extremely low values taken by $r_\mathrm{\textsc{f}}^d$ indicate that the measured flux of muons is mainly composed by upward going muons coming from backward directions. Conversely, the backward ratio $r_\mathrm{\textsc{b}}^d = \phi_\mathrm{\textsc{b}}^d / \phi_\mathrm{\textsc{f}}^u$ (blue curve) does not significantly depart from $1$ excepted at slopes just above the horizon where $r_\mathrm{\textsc{b}}^d = 0.9$. This indicates that almost no upward flux $\phi_\mathrm{\textsc{f}}^u$ is present in the measured flux. This agrees with the strong obstruction in the forward direction that efficiently stops eventual upward going muons.

The $r^d$ curves obtained for the \textsc{etomo} data set (rightmost top plot of Fig. \ref{FluxRatio}) look very similar to the \textsc{smtomo} curves and reflect the obstruction asymmetry observed for this location (Fig. \ref{AcquisitionSites}). However, $r^d$ values significantly lower than $1$ are obtained in the backward direction for $\alpha_\mathrm{\textsc{b}} > -5^{\circ}$, in agreement with the moderate forward obstruction for the corresponding range of slopes.

The $r^d$ curves obtained for \textsc{smtomo} and \textsc{etomo} indicate the presence of a strong backward upward flux, $\phi_\mathrm{\textsc{b}}^u$, in the forward flux measured in a slope range of a several degrees above the horizon ($3^{\circ}$ for \textsc{smtomo} and $4^{\circ}$ for \textsc{etomo}). This strong flux is not detected for the \textsc{rfcalib} data set probably because of the lack of data around the horizon due to the horizontal position of the telescope. At larger slopes, the upward flux is weaker and appears quite stable. On \textsc{smtomo} we recover the $0.2~\mathrm{sr}^{-1}.\mathrm{cm}^{-2}.\mathrm{day}^{-1}$ of \textsc{rfcalib}. This stable zone is less obvious on \textsc{etomo}. The forward upward flux takes significant value in the "strong zone", lower than its equivalent forward upward flux as if the volcano had partially absorbed it. Below, the noise is too important for any interpretation but it can fit with the stable backward zone. The particle flux for the \textsc{etomo} data set is about four times stronger than for \textsc{smtomo}. This may be due to the difference of altitude of the two locations (Table~\ref{TableSites}).

\section{Remarks about the origin of upward-going muons}

The results presented in section \ref{DataAnalysis} clearly establish the existence of a flux of upward-going particles when the rock obstruction below the horizontal is less than several tens of meters. The \textsc{lbcalib} data acquired at a location surrounded by wide horizontal flat area is used to estimate the noise level and establish the significance threshold for the other data sets analysed in the present study. A tiny upward flux seems to exist for the \textsc{lbcalib} data in a small range of slopes just below the horizontal. However, the low signal-to-noise ratio of this data set makes the existence of this flux questionable.

For the data sets acquired on La Soufri\`ere and Mount Etna, the origin of upward-going muons detected just beneath the horizontal may be explained by accounting for the altitude of the measurement locations with respect to the nearby sea surface. At such altitudes, sub-horizontal atmospheric showers whose axis has an apparent negative slope in the telescope frame can produce muons able to reach the detectors without any obstruction along their path. For the altitude of the experiments considered in the present study, the maximum slope equals $1.7^{\circ}$ for Mount Etna and $1.2^{\circ}$ for La Soufri\`ere.

For muons produced by showers coming from below the horizontal atmospheric slant depth, $\varrho_z$, is increased by the distance-to-horizon length, $L_z$,
\begin{equation}
\varrho_z = \varrho_0 + L_z \times \overline{\rho}
\end{equation}
where $\overline{\rho}$ is the atmosphere average density and $\varrho_0 = 36000~\mathrm{g}.\mathrm{cm}^{-2}$ is the horizontal slant depth at $z=0$. For the measurement sites considered in the present study, $\varrho_z \approx 55000~\mathrm{g}.\mathrm{cm}^{-2}$, a value noticeably larger than the horizontal slant depth $\varrho_0$ usually considered as an upper limit in shower models. A cosmic muon has to spend a supplementary energy loss of $\delta E \approx 40~\mathrm{GeV}$ in addition to the energy loss $\delta E_0 \approx 70~\mathrm{GeV}$ corresponding to $\varrho_0$. Consequently, only muons with an initial energy $E \apprge 110~\mathrm{GeV}$ are able to reach the telescope. 

The mechanism invoked above and involving sub-horizontal showers cannot explain the existence of the flux of upward-going muons observed at large negative slopes in the \textsc{smtomo} and \textsc{etomo} data sets. Indeed, the slopes involved ($\alpha = 5^\circ - 15^\circ$) are too important to allow muons created high in the atmosphere to reach the telescope, and one must resort to muons created near the ground in the volume of atmosphere located below the altitude of the telescope. The main part of the muon flux observed at the ground level comes from hadron atmospheric showers where charged pions are the source of muons through $\pi^{+ (-)} \rightarrow \mu^{+ (-)} \,+\, \nu_\mu (\bar{\nu}_\mu)$ decay occurring high in the atmosphere. In such showers, momentum is transferred from the primary particle to the shower particles whose trajectory only slightly depart from the shower axis, and hadron shower may then hardly be the source of muons observed in our experiments. At lower energy instead (e.g. decay of rest pions) the muon angle with respect to the decaying particle may be large, and a weak pion located nearby the telescope can produce an upward muon if it decays below the instrument. This agrees with the observed obstruction dependence since these low-energy muons are stopped by a little obstruction. It explains the stable zone of the flux versus the zenith angle as the pion disintegration is isotropic. Gamma-induced electromagnetic showers may also produce a tiny flux of muons either by photoproduction of pions followed by decay into muons and neutrinos as in hadron showers or by direct pair production.

It must be kept in mind that other particles than muons may also contribute to the detected upward flux. Electrons and charged pions are unlikely to be detected in a tomography configuration because pions flux is very weak and electron cross section is very high as compared to the muon. However if there is no obstruction like for the backward flux of the tomography acquisitions and the forward flux of \textsc{rfcalib} and even if the obstruction is little like for the \textsc{lbcalib} acquisition and the backward flux of \textsc{rfcalib} (fig. \ref{AcquisitionSites}) they may have a little impact. These scenarios need further dedicated Monte Carlo simulations. Indeed if the cosmic particles flux at sea level for low zenithal angles is very well known (Barret et al 1952; Golden et al. 1995; Grieder 2001; Hebbeker 2002), only very specific cases have been studied for horizontal particles (Ave et al. 2000a, Ave et al. 2000b). Our simulations will be discussed in a forthcoming paper. 

\section{Influence of upward noise on density radiographies}

We now examine the influence of the flux of upward-going muons on density radiographies, and give an example for La Soufri\`ere of Guadeloupe. Basically, a density radiography is obtained by computing the opacity, $\varrho \; [\mathrm{hg.cm}^{-2}]$, for each line of sight spanned by the telescope:
\begin{equation}
\varrho (L) = \int_{L} \rho(\xi) \mathrm{d}\xi = \overline{\rho} \times L,
\label{OpacityDefinition}
\end{equation}
where $\xi$ is the coordinate measured along the ray trajectory of length $L$ across the volcano of density $\rho$. Once determined, the opacity is converted into the average density $\overline{\rho} = \varrho / L$ to construct the radiography image.

In practice, the opacity value is determined by searching the cut-off energy $E_{\mathrm{min}}(\varrho)$ that reproduces the measured flux of muons by integrating the incident differential flux $\Phi_{0} \; [\mathrm{cm}^{-2}\,\mathrm{sr}^{-1}\,\mathrm{s}^{-1}\,\mathrm{GeV}^{-1}]$,
\begin{equation}
\phi = \int_{E_{\mathrm{min}}(\varrho)}^{\infty} \Phi_{0}(E,\gamma)\mathrm{d}E \;\; [\mathrm{cm}^{-2} \mathrm{sr}^{-1} \mathrm{s}^{-1}].
\label{IntegratedSpectrum}
\end{equation}
A discussion concerning the models available for $\Phi_{0}$ may be found in Lesparre \textit{et al.} (2010) and, in the present study, we use the model given by Tang \textit{et al.} (2006). Equation (\ref{IntegratedSpectrum}) shows that a positive bias in the measured flux $\phi$ conducts to an underestimate of $E_{\mathrm{min}}$ hence of opacity $\varrho$.

To document the sensitivity of the reconstructed opacity to the flux of upward-going muons, we apply the upward flux correction to the \textsc{smtomo} dataset on the forward downward flux. Computing $r^d$ on each sight axis of the telescope, we use the results of fig. \ref{FluxRatio} and fit a simple polynomial law on it. Finally we apply eq. \ref{eqn_ratio4}. One can see on the obstruction figure of \textsc{smtomo} the limits of this approach (fig. \ref{AcquisitionSites}). Indeed the site has a significant backward obstruction on the centre of the acquisition window on the first $4^{\circ}$ below the horizon whereas we find open-space on the borders.

\begin{figure}
\begin{center}
\includegraphics[width=\linewidth]{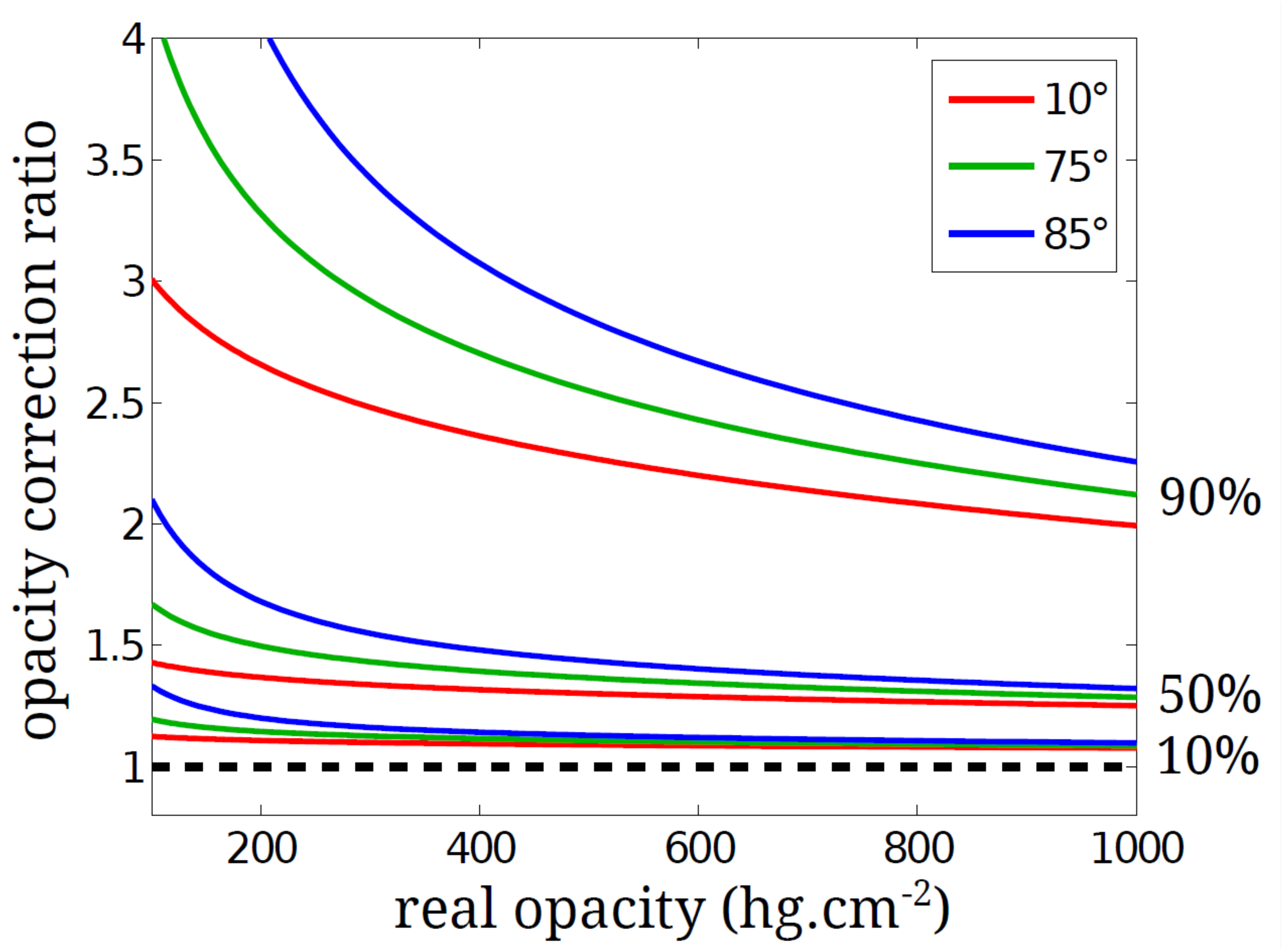}
\vspace{-8mm}
\end{center}
\caption{Impact of an error measured on the flux on the obstruction. We use Tang model at sea level. For example if we measure an opacity of $800~\mathrm{g}.\mathrm{cm}^{-2}$ at a slope of $15^{\circ}$ and have a $50\%$ downward ratio, the real opacity is equal to $800~\mathrm{g}.\mathrm{cm}^{-2}$ times its corresponding opacity correction factor (here $1.5$).}
\label{modelErrorImpact}
\end{figure}

Let us consider the impact of the correction of the strong upward flux on the first degrees below the horizon. In that case the obstruction is about $1000~\mathrm{m}$, and as the density is expected to be around $1.5~\mathrm{g.cm}^{-3}$ the opacity reads $1500~\mathrm{hg.cm}^{-2}$. According to Fig. \ref{modelErrorImpact}, it corresponds to a significant correction ratio of about 1.5. The long range and weak upward flux provides a $0~\%$ to $80~\%$ correction for obstructions going from $200~\mathrm{m}$ to $1000~\mathrm{m}$ (opacities going from $300~\mathrm{hg.cm}^{-2}$ to $1500~\mathrm{hg.cm}^{-2}$. So progressively going from $20^{\circ}$ above to horizon to the horizontal, we get a correction ratio going from $1$ to $1.3$.

\begin{figure}
\begin{center}
\includegraphics[width=\linewidth]{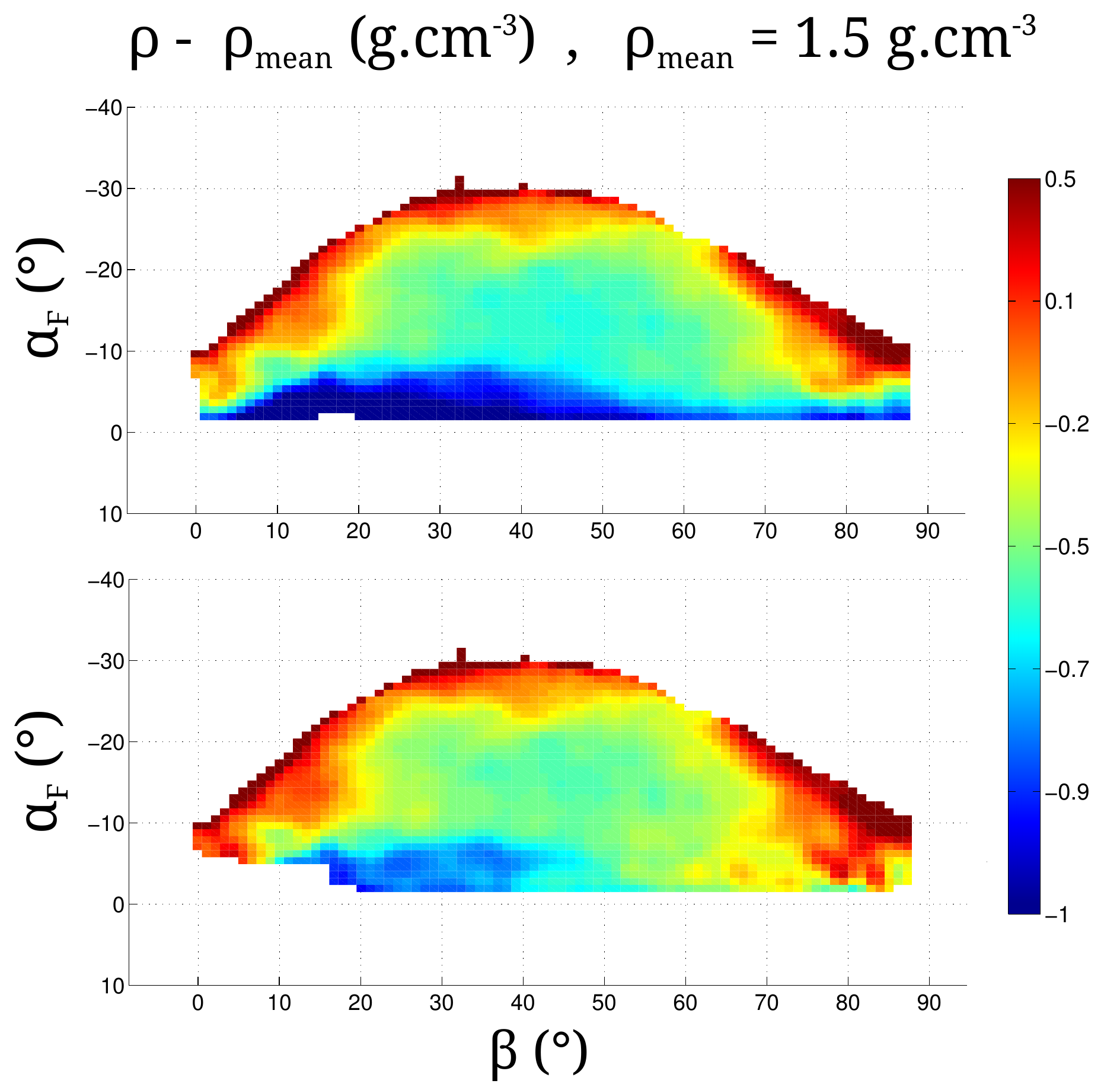}
\vspace{-8mm}
\end{center}
\caption{Tomography result of \textsc{smtomo} high definition acquisition without upward flux correction (TOP) and with upward flux correction (BOTTOM). Rock density in $\mathrm{g}.\mathrm{cm}^{-3}$.}
\label{density_SAMHD}
\end{figure}

Top part of fig. \ref{density_SAMHD} shows the raw results of the tomography. Between the $\alpha_F = -10^{\circ}$ and the horizon we observe a progressive decrease of the density, down to $0.5~\mathrm{g.cm}^{-3}$, which is obviously wrong. On the left side of the window, for $\beta$ taking values between $5^{\circ}$ and $50^{\circ}$ the decrease is way quicker and is correlated with an obstruction significantly thicker (fig. \ref{AcquisitionSites}). 

Bottom part of fig. \ref{density_SAMHD} shows the corrected tomography results. On the left side of the acquisition we did not delete the very low density zone previously mentioned. The density is significantly higher but still lower than on the rest of the picture. On the right side of the acquisition window (for $\beta$ taking values between $50^{\circ}$ and $85^{\circ}$) the correction appears if efficient. The density fluctuations are smoother and we suppressed the very low values zone over the horizon to agree with the rest of the acquisition. Finally for $\alpha_F < -10^{\circ}$ the correction provided is quite insignificant. 


\section{Conclusion}

The data analysis discussed in the present study demonstrates the existence of a flux of upward-going muons whose trajectories might be confounded with those of downward-going muons crossing the volcano to radiography. Upward fluxes are detected only when the back side of the telescope is exposed to a wide the volume of atmosphere located below the measurement level. This favours the existence of processes like photo-production of pions or of pairs of muons at low altitude, near the ground level. High-resolution clocking systems (fig. \ref{fig:tdc} ) are mandatory to separate the upward and downward fluxes (fig. \ref{telescopeFlyingTimeHistosVsZenithAngle}).

In some instances, the upward flux can be as intense as the sought downward flux, leading to important bias in the reconstructed opacity of the volcano (fig. \ref{modelErrorImpact}). De-biasing based on statistical considerations may be applied to retrieve the correct opacity (fig. \ref{density_SAMHD}). Such corrections are indispensable to perform accurate density 3D tomography of volcanoes that need to combine radiographies acquired at different view angles and subject to different intensity of upward flux.

\begin{acknowledgments}
Field operations in Guadeloupe received the help from colleagues of the Volcano Observatory, from the crews of the helicopter station of the French Civil Security (\url{www.helicodragon.com}) and from members of the National Natural Park of Guadeloupe (\url{www.guadeloupe-parcnational.fr}). On-field maintenance and servicing of the telescope are ensured by Fabrice Dufour. Field operations on Mount Etna received the help of colleagues of the Volcano Observatory at Catania. Logistic organization was ensured by the \textsc{ulisse-in2p3} department of \textsc{cnrs} (\url{ulisse.cnrs.fr}). We acknowledge the financial support from the UnivEarthS Labex program of Sorbonne Paris Cit\'e (\textsc{anr-10-labx-0023} and \textsc{anr-11-idex-0005-02}). This is IPGP contribution ****.
\end{acknowledgments}

\label{lastpage}
\end{document}